\documentclass[a4paper]{jpconf}
\usepackage{graphicx}
\def\be{\begin{eqnarray}&&}
\def \nonu {\nonumber \\&&}
\def \lfp { {\downarrow}}
\def\psla{\slash \! \! \!}

 \def\ee{\end{eqnarray}}
\begin{document}
\title{Light-front projections of the  Bethe-Salpeter amplitude and the
4D electromagnetic current  for an interacting two-fermion system}

\author{T Frederico$^1$, J A O Marinho $^{1,2}$,
E Pace$^3$ and
G Salm\`e$^4$ } 
\address{ $^1$ Dep. de F\'\i
sica, Instituto Tecnol\'ogico de Aeron\'autica,  12.228-900 S\~ao
Jos\'e dos
Campos, S\~ao Paulo, Brazil\\
$^2$Centro de F\'\i sica Computacional,
Department of Physics,
University of Coimbra, Portugal\\
$^3$ Dipartimento di Fisica, Universit\`a di Roma "Tor Vergata"
and Istituto Nazionale di Fisica Nucleare, Sezione Tor Vergata,
Via della Ricerca
Scientifica 1, I-00133  Roma, Italy \\
$^4$Istituto  Nazionale di Fisica Nucleare, Sezione Roma I, P.le A.
Moro 2, I-00185 Roma, Italy }

\ead{salmeg@roma1.infn.it}

\begin{abstract}
A recent approach for constructing an exact 3D reduction of the 4D
matrix elements of the  electromagnetic current for an interacting two-fermion
 system, is briefly reviewed. The properties of the obtained 3D current,
like the fulfillment of the Ward-Takashi Identity and its Fock decomposition,
will be illustrated in relation with  future applications to few-nucleon
systems. 
\end{abstract}

\section{Introduction}

A detailed investigation of the electromagnetic (em) properties  of
hadrons and  nuclei requests a careful treatment of the
current operator for interacting fermion systems. In particular, besides general
properties like the Poincar\'e covariance,
 the gauge symmetry of the em interaction plays the well-known essential role, that leads to 
the Ward-Takahashi Identity (WTI) for the em vertex function. 
In this short presentation, some insights will be
given on the main ingredients of our approach \cite{MFPSS} aimed to construct a 3D
em current for an interacting two-fermion system, that i) fulfills WTI and
therefore allows the current conservation (CC) and ii) preserves at large extent
the Poincar\'e covariance, emphasizing the relevant steps for obtaining
workable expressions to be used in actual calculations.

Field theoretical approaches, based on the Bethe-Salpeter
equation, (see, e.g. \cite{sales00,sales01,adnei07}) allow one
 to evaluate matrix elements of the 4D em current for an interacting 
 two-body system, viz
\be
\langle P_f, J_f,\sigma_f|j^\mu|P_i,J_i,\sigma_i\rangle =
{\cal C}\sum_{\tau_i,..}\int {d^4k_1 \over (2\pi)^4} 
\bar \Psi^{J_f}_{ \sigma_f}(k_1,k^\prime_2,P_f)
{\cal J}^\mu(k_1,k^\prime_2,k_2)
 \Psi^{J_i}_{\sigma_i}(k_1,k_2,P_i)
 \label{me4d}
\ee
where, in the present case,  $\Psi^{J \sigma}(k_1,k_2,P)$ is the Bethe-Salpeter (BS) amplitude of a
two-fermion systems (with $k_1+k^\prime_2=P_f$ and $k_1+k_2=P_i$ for the final
and initial state, respectively),  ${\cal C}$ contains the normalization
factors, and 
${\cal J}^\mu$ is the 4D
fermion-photon vertex (or kernel) that contains one + two +...+ many-body 
contributions, and must 
fulfill WTI. The sum in Eq. (\ref{me4d}) is performed on isospin, spinorial, etc.
indexes.

To solve the BS equation in  Minkowski space for obtaining 
$\Psi^{J \sigma}(k_1,k_2,P)$, in realistic cases, is still a
prohibitive task, but new and very effective approaches are coming into play
(see, e.g. \cite{carb}). 
On the other hand, 3D reductions are well known tools (see, e.g.
\cite{grossbook}) 
for attempting to
overcome the above mentioned difficulties, but still saving at large extent the
general properties that the em current has to fulfill.

Aim of our investigation \cite{MFPSS} is i) to achieve an exact 3D reduction of the
4D matrix elements in Eq. (\ref{me4d}), i.e. without loosing any physics
content, and  ii) to elaborate a systematic analysis
of the obtained expression in order to develop effective approximation.
The main ingredients for our  approach, that were already applied to the case of
two-boson interacting system in \cite{adnei07}, are in order i) the 
Quasi-Potential (QP) approach for the
Transition matrix (see \cite{wolja}), and ii) the projection of 
the 4D physical quantities 
onto the 3D
Light-front (LF) hyperplane (i.e. $x^+=x^0+x^3=0$) 
(see, e.g. \cite{sales00,sales01}). As a first result,  one can establish a 
formally exact
correspondence between
 4D BS amplitude and the 3D LF "valence" wave
function. Then it is possible to express, without approximation, the 
matrix elements of the 4D em current, where the BS amplitudes appear, in terms
of matrix elements  of  a 3D LF current, that fulfills the
WTI. It should be pointed out that in these last matrix elements the valence
states are present.

Within the Quasi-Potential approach
 for the T-matrix, where the so-called {\em effective
 interaction} appears, one can develop a meaningful approximation scheme 
 for evaluating the 3D matrix elements, based on the Fock decomposition of the
 effective interaction. In particular, one expects a   
 decreasing
influence  of the
 Fock states that contain an  increasing number of  intermediate particles. This machinery 
  allows one to construct an  iterated  approximation of the effective
 interaction and eventually to evaluate matrix elements of the "truncated" 
 LF em current,
  that still
 fulfills at any order the LF WTI.
 Finally, by using a Yukawa model in ladder approximation, as a pedagogical example, 
 explicit
 expressions for  many-body contributions  to the em LF current have been obtained
 \cite{MFPSS}.

\section{BS amplitudes and EM current} 
The quantities present in Eq. (\ref{me4d}) are the BS amplitude and the em
vertex. In particular, the BS amplitude $| \Psi \rangle$,   that depends 
upon {\em internal
variables}, satisfies the following equation (the $\lim_{\varepsilon\to 0}$ is understood)
\be G^{-1}(K)\left| \Psi \right\rangle =0~ \label{BSE}\ee with
appropriate boundary conditions for bound and scattering
states. In Eq. (\ref{BSE}), $K$ is the  4-momentum of the center of mass and  
 the inverse of the Green function is given by 
\be  G^{-1}(K)=G^{-1}_0(K)-V(K)\label{green}\ee
where  $V(K)$ is the 4D interaction, as obtained by the Lagrangian, 
 $G_0(K)=S(k_1)~S(K-k_1)$, with   $ S(k)$ 
the Dirac propagator. 

  The em vertex, ${\mathcal J}^\mu$, corresponding to the
interaction $V(K)$ has a free term, ${\mathcal J}_0^\mu$, and
another one, ${\mathcal J}_I^\mu$, which depends on the interaction,
as requested by the commutation rules with
the generators of the Poincar\`e group. 
 It also
 satisfies the following Ward-Takahashi
identity~  \be Q_\mu{\mathcal
J}^\mu(Q)=G^{-1}(K_f)\widehat e~-\widehat e~G^{-1}(K_i)
\label{wti}\ee where $Q^\mu$ is the 4-momentum transfer 
and $\widehat e=\widehat e_1+\widehat e_2$ is the pointlike charge operator 
  with matrix elements given by
$\langle k_j|\widehat
e_j|p_j\rangle=e_j\delta^4\left(k_j-p_j-Q\right)$

 Combining Eqs. (\ref{green}) and (\ref{wti}) with the decomposition of 
 ${\mathcal J}^\mu$ in terms of the dependence upon $V(K)$, it turns out that
 each contribution fulfills a suitable constraint, viz
$$
Q_\mu{\mathcal J}_0^\mu(Q)= G^{-1}_0(K_f)\widehat e~-\widehat
e~G^{-1}_0(K_i)  $$ and $$
Q_\mu{\mathcal J}_I^\mu(Q)= \widehat e~ V(K_i)-V(K_f)\widehat e~
.$$ 
Finally,  
 CC can be explicitly retrieved by using solutions of the 
BS equation, Eq. (\ref{BSE}),
i.e.,  
 \be Q_\mu\left\langle \Psi_{f}\right|{\mathcal J}^\mu(Q)\left|
\Psi_{i}\right\rangle=\langle \Psi_{f}|~\left[G^{-1}(K_f) \widehat
e~-\widehat e~G^{-1}(K_i)\right]~|\Psi_{i}\rangle =0~~~.\ee 

 After introducing the general equations fulfilled by BS amplitudes and em
 current, let us illustrate our procedure. As well known, the T-matrix is 
 given in terms of the interaction $V(K)$ by an
integral equation: 
$ T(K)=V(K)+V(K)~G_0(K)~T(K)$. 
 
In the  Quasi-Potential framework \cite{wolja},
$T(K)$ can be rewritten in terms of two coupled equations
 \be
T(K)=W(K)+W(K)~G_{aux}(K)~T(K)
\ee
where  the effective interaction  $W(K)$,  in turn,
is a solution of
 \be
W(K)=V(K)+V(K)~\Delta_0(K)~W(K)
\label{qp2}\ee 
with $\Delta_0(K):=G_{0}(K)-G_{aux}(K)$ , and   $G_{aux}(K)$ is a smart
choice to speed up the convergence of the iterative solutions.

\section{Projecting onto the Light-Front hyperplane}  
In Eq. (\ref{me4d}) the 4D integration contains all the difficulty
 given by the very complicated analytic structure of
both BS amplitudes and em vertex in Minkowski space. The
adopted strategy  for decreasing the degree of difficulty, is based on
disentangling the "dynamical" variable, i.e. the time, from the remaining three
kinematical variables, that describe the "initial" state. In order to accomplish
such a task, the LF kinematics appears a very appealing tool (see, e.g.
\cite{brodsky} for an extensive review) since the  LF combination of the
Poincar\'e generators has the property to produce the maximal numbers of
generators without interaction (7 out of 10). In turn, such a property produces
many other relevant features, like the kinematical nature of the LF boosts (to be
not confused with the standard boosts, a part the longitudinal one).
In particular, the LF components of the 4-momentum  and the scalar product
are given by 
\be k^{\pm}=k^0\pm k_z, \quad  \quad {\bf k}_\perp\equiv \{k_x,k_y\}, 
\quad \quad k\cdot x={(x^-k^++ x^+k^-)\over 2} -{\bf x}_\perp 
\cdot{\bf k}_\perp~~~.\ee
 Integrating a given physical quantity, $\phi(k)$, over the minus component 
of the 4-momentum, $k^-$, amounts to restrict its
dependence upon only 3 variables in the coordinate space. One has
\be
\int dk^- ~ \phi(k)=\int dk^- ~{1 \over 2 (2 \pi)^4}\int dx^+ ~dx^-~
d{\bf x}_\perp ~e^{i k \cdot x}~ \tilde \phi(x)=\nonu=
{1 \over  (2 \pi)^3}\int dx^+ ~dx^-~ \delta(x^+)
d{\bf x}_\perp ~e^{i(x^-k^+/2 -{\bf x}_\perp \cdot{\bf k}_\perp )}~ 
\tilde \phi(x) ~~~.
\label{proj1}\ee 
Therefore, the "LF-time", $x^+$, that
labels the dynamical evolution (as the standard "t" in the instant form of the
relativistic Hamiltonian), is constrained to its initial value, i.e.
$x^+=0$.  Therefore, once the physical quantities are integrated 
over $k^-$, they
are constrained  to live onto the LF hyperplane defined by $x^+=0$.

 A crucial step for the LF
projection is the separation of the fermion propagator in an
on-shell term and in an instantaneous one, i.e
\be i~S(k)=
\frac{\psla{k}+m}{k^2-m^2+i\varepsilon}=
\frac{\psla{k}_{on}+m}{k^+(k^--k^-_{on} +{i\varepsilon\over
k^+})}+\frac{\gamma^+}{2 k^+} = i~S_{on}(k)+\frac{\gamma^+}{2 k^+}\label{proj2} \ee
where $k^-_{on}=({\bf k}_{\perp}^2+m^2)/k^+$ is the minus-component of
$k^\mu_{on}$, such that $k_{on}\cdot k_{on}=m^2$,
 and  the second term   leads to an instantaneous (in LF time!)
free propagation, since the formal  Fourier transform generates 
 $\delta(x^+)$ (see Eq. (\ref{proj1})).  
 It should be pointed out that such a  term makes the treatment of a  fermionic 
system basically different from the treatment of a  bosonic  one. On the other
hand, the decomposition in Eq. (\ref{proj2}) suggests a strategy for achieving our
goal of an exact 3D reduction. As a matter of fact, let us define 
the 4D on-shell two-body Green's function \cite{MFPSS,adnei07}
 \be{G}^{on}_{0}(K)  := S_{on}(k_1)~S_{on}(K-k_1) \label{proj3}\ee  
that plays a key role in our approach, since
allows us to separate the  {\em trivial}  4D propagation associated to such a
contribution. 

The 3D LF counterpart of ${G}^{on}_{0}(K)$ is
introduced by integrating over   $k^-_1$ and  $k^{\prime -}_1$,
 i.e. by projecting   
 onto the 3D LF hyperplane, the following matrix elements (note that the 
 other
 three components of the momentum remain operators)
 \be g_{0}(K)=~\lfp {G}^{on}_{0}(K)
\lfp~\equiv \Omega^+ \int dk_{1}^{\prime
-}dk_{1}^{-}\left\langle k_{1}^{\prime -}| {G}^{on}_{0}(K)|
k_{1}^{-}\right\rangle \ee  
where $\Omega^+$ is a suitable kinematical factor ensuring the kinematical
Poincar\'e covariance (e.g. $\Omega^+= k^+$ or  $\Omega^+= \sqrt{k^+(K^+-k^+)}$), and the symbol $\lfp$, from
now on, indicates the integration over  the minus component
of the 4-momentum that appears on the left or on the right side, depending upon the
position of the symbol itself.

In the case without interaction, it is easy to find the relation between 
the 3D LF   free state, $|\phi_0\rangle$,  solution of
$ g^{-1}_0(K)~|\phi_0\rangle =0$ 
(note that instantaneous  terms do not contribute)
and  the free BS  amplitude,  solution of
$ G^{-1}_0(K)~|\Psi_0\rangle =0$ (note that the particles are on their own
mass-shell in the free case).  It turns out \cite{MFPSS} that 
\be|\Psi_0\rangle = G^{on}_0(K)\lfp ~g^{-1}_0(K)~
|\phi_0\rangle=\Pi_0(K)~|\phi_0\rangle\ee 
where  $\Pi_0(K)$ is the {\em free reverse LF projector}, that allows to jump
from a 3D description to a 4D world. The operator   $\Pi_0(K)$ has  
a simple explicit
expression \cite{MFPSS,adnei07} in terms of the positive
energy Dirac projector.     
Let us remind that, by construction
 $ \lfp  |\Psi_0\rangle =|\phi_0\rangle$: this completes the reconstruction
 pattern in the simple case of a free two-fermion system.

 In order to apply the projection technique to the QP approach, 
 namely for an interacting case, one
 needs an intermediate step, given by a suitable choice for the auxiliary
 Green's function to be used in the 4D operator $\Delta_0$, see Eq. (\ref{qp2}).
 The chosen 4D auxiliary Green's function is defined as follows in terms of the 3D $g_0(K)$
 \be G_{aux}(k)= \Pi _0(K) g_0(K) \overline\Pi_0(K) \ee
 where $\overline\Pi_0(K)=g^{-1}_0(K)~\lfp G^{on}_0(K)$.
Then, the  T-matrix  and the effective interaction $W(K)$
 can be rewritten as follows 
\be
T(K)=W(K)+W(K)~\left [\Pi _0(K) g_0(K) \overline\Pi_0(K)\right]~T(K)
\nonu
W(K)=V(K)+V(K)~\Delta_0(K)~W(K)=\nonu=V(K)+V(K)~
\left[G_{0}(K)-\left(\Pi _0(K) g_0(K) \overline\Pi_0(K)
\right)\right]~W(K)~~~.
\ee
 The equation for $W(K)$
can be solved by iteration, i.e.
 \be   W(K)=V(K)~\sum_{i=1}^\infty \left[\Delta_0(K)V(K)\right]^{i-1}~~~.
 \label{iter}\ee
 The  rate of convergence of this series is related to
$\Delta_0(K)$,
that describes the difference between the intermediate propagation of two
non-interacting fermions and the 4D counterpart of the  LF propagator
$g_0(K)$. Depending upon the powers of $\Delta_0(K)$, we will have intermediate
states with more and more bosons and fermions (once we introduce  time-ordered diagrams).

 The Rosetta Stone relating 4D and 3D quantities, is readily 
constructed. The 3D T-matrix and effective interaction are  respectively 
\be t(K)=\overline\Pi_0(K)~T(K)~\Pi _0(K)  \quad \quad \quad
w(K)=\overline\Pi_0(K)~W(K)~\Pi _0(K)~~~.\ee

Once $t(K)$ is reexpressed in terms of $g_0(K)$ and $w(K)$,  one can
introduce the 3D
 interacting Green's function  and its inverse as follows 
  \be
 g(K)=g_0(K)+g_0(K)t(K)g_0(K) \quad \quad \quad g^{-1}(K )
 ={\overline \Pi}_{0}(K)~G^{-1}(K) ~\Pi(K)\ee 
where the {\em interacting LF
 reverse projection operator} is
\be \Pi(K)=~\left [1 +\Delta_0(K)~W(K)~\right ]~ \Pi_{0}(K)~~~.\ee 
  
  The solution $\left| \phi \right\rangle$ of the following 3D equation,
\be
g^{-1}(K )\left| \phi \right\rangle = \left[g^{-1}_0(K
)-w(K)\right]~\left| \phi \right\rangle=0 \ee
with appropriate boundary conditions for bound and scattering
states, is  the valence component  of the LF wave
function for two interacting fermions, as can be deduced by the relation with
the BS amplitude, shown in what follows.
The exact relations between the 3D
valence component and the 4D BS amplitude can be readily obtained, and they are
given by
 \be\left|
\Psi \right\rangle=\Pi(K)~\left| \phi
\right\rangle \quad \quad \quad \lfp  G^{on}_0(K) ~G^{-1}_0(K)~
\left| \Psi
\right\rangle=~ \left| \phi \right\rangle ~~~. \label{3d4d}\ee
The rightmost equation illustrates the motivation for calling
 $\left| \phi \right\rangle$ valence
component, given the presence of the integration on the minus component in lhs
(see, e.g \cite{carb} and references
quoted therein). Such an integration makes  only the first component
of the Fock decomposition of $\left| \Psi
\right\rangle$ to survive.

A possible approximation scheme is suggested by the
QP expansion of the effective interaction, Eq. (\ref{iter}). 
Then, $w(K)$, $t(K)$ and  the 3D LF Green's
function $g(K)$ (all  containing $W(K)$ in a more or less explicit way) can be evaluated 
by truncating the expansion at a given order.
It should be pointed out that the full
complexity of the Fock-space affects $g(K)$ through the effective
interaction $w(K)$, as can be seen through a 
time-ordered analysis of the diagrams associated to each contribution 
$\left[\Delta_0(K)V(K)\right]^{i-1}$, and reminding that $\Delta_0(K)$ 
describes propagation.   Indeed, the truncation of the QP expansion
 limits the number of Fock states involved in
the construction of the effective interaction.   
As a final remark, 
 one could argue that
the convergence rate of the QP expansion is related to the small
 probability of the higher Fock-components.

\section{The 3D EM current and  the LF Ward-Takahashi Identity} 
Once we have obtained the relation between the 4D BS amplitude and
the 3D LF valence wave function, one can establish a direct link between 
the matrix
element of the 4D current and the matrix elements of the 3D LF em current 
operator. 

For both scattering and bound states one has
\cite{MFPSS,adnei07,sales00,sales01}
\be
\left \langle\Psi_{f}\right|{\mathcal J}^\mu(Q)\left|
\Psi_{i}\right\rangle=\langle\phi_{f} |j^\mu(K_f,K_i) |\phi_i\rangle
\ee
where  the  3D LF current operator, acting on
the valence wave functions, is defined as follows
\be 
j^\mu(K_f,K_i)=  \overline \Pi(K_f){\mathcal J}^\mu(Q)\Pi(K_i)=\nonu =
\overline\Pi_0(K_f)\left[1+
W(K_f)\Delta_0(K_f)\right]{\mathcal
J}^\mu(Q)\left[1+\Delta_0(K_i)W(K_i)\right]
\Pi_{0}(K_i)\label{j3d}~~~.\ee

Since the 4D current must fulfill the WTI, Eq. (\ref{wti}), the 3D 
LF current fulfills the  following LF WTI \cite{MFPSS}
\be Q_\mu j^\mu(K_f,K_i) = \overline \Pi(K_f)\left[G^{-1}(K_f)\widehat
e-\widehat e G^{-1}(K_i)\right]\Pi(K_i)=\nonu =
g^{-1}(K_f)~\widehat {\cal Q}^L_{LF}- \widehat {\cal
Q}^R_{LF}~ g^{-1}(K_i) 
\label{wti3d}\ee
where  ${\cal Q}^{L(R)}_{LF}$ is called the left (right) LF 
charge
operator (note that it is {\em interaction free !}). For the sake of
concreteness, let us show the explicit expression of the left charge, viz
\be \widehat {\mathcal Q}^L_{1LF}= \Lambda_+(\widehat
k_{1on})\frac{m_1}{\widehat k^{+}_1}\gamma_1^+~\widehat e_{1LF}
\Lambda_+(\widehat k_{1on})\Lambda_+(\widehat k_{2on})\ee
where $\Lambda_+$ is the positive energy Dirac
projector, and $\widehat e_{1LF}$  the 3D pointlike charge operator, given by
\be
\langle k^{\prime+}_1,{\vec k}^\prime_{1\perp}|\widehat
e_{1LF}|k_1^+,\vec
k_{1j\perp}\rangle:=e_1\delta\left(k^{\prime+}_1-k^+_1-Q^+\right)~
\delta^2\left({\vec
k}^\prime_{1\perp}-\vec k_{1\perp}-\vec Q_{\perp}\right)~~. \label{d3}
\ee
The current conservation straightforwardly follows by
taking the matrix elements of Eq. (\ref{wti3d}) between   3D interacting
states $|\phi_{i(f)}\rangle$, solutions of $g^{-1}(K)~|\phi\rangle =0$

For the actual calculations, one needs to truncate the iterated expression of
the effective interaction $W(K)$, Eq. (\ref{iter}), and therefore one ends up
with a 
truncated 3D em current.
The truncation scheme for the LF current cannot be carried out in a naive way, if
one has to fulfill a WTI. Indeed, at each truncation order it will correspond a
truncated effective interaction, $w^{(n)}(K)$, a truncated LF Green's function $g^{(n)}(K)$, a truncated valence wave function, 
$|\phi^{(n)}\rangle $, a properly truncated 3D
LF current, and finally a truncated WTI. Let us see in details this chain.

Once we truncate the QP expansion of the effective interaction at the order $n$,
\be W^{(n)}(K)=V(K)\sum_{i=1}^n \left[\Delta_0(K) V(K)\right ]^{i-1}\ee 
we can immediately generate an approximate 3D LF Green's function, 
$g^{(n)}(K)$, leading to an 
approximate valence wave functions, $|\phi^{(n)}\rangle$, solutions of the
following equation
\be [g^{(n)}(K)]^{-1}~|\phi^{(n)}\rangle=\left [g_0^{-1}(K)-w^{(n)}(K) \right
]~|\phi^{(n)}\rangle=0~~~.\ee
  The corresponding em current, that fulfills   CC, if one uses 
  $|\phi^{(n)}\rangle$ for evaluating the matrix elements, cannot be directly 
 constructed inserting in the definition of the 3D current, Eq. (\ref{j3d}),
  the truncated 
 $W^{(n)}(K)$, i.e.
\be Q\cdot j^{(n)}(K_f,K_i) \ne 
[g^{(n)}(K_f)]^{-1}~\widehat {\cal Q}^L_{LF}- \widehat {\cal
Q}^R_{LF}~[g^{(n)}(K_i)]^{-1}\ee
with 
\be j^{(n)\mu}(K_f,K_i)=\overline\Pi_0(K_f)\left[1+
W^{(n)}(K_f)\Delta_0(K_f)\right]{\mathcal
J}^\mu(Q)~\times \nonu\left[1+\Delta_0(K_i)W^{(n)}(K_i)\right]
\Pi_{0}(K_i)~~~. \ee
 The fulfillment of the LF WTI excludes such a naive approximation scheme for
 the truncated current, but, at the same time, 
gives us the correct
hint. One should implement a  power counting of the effective interaction that
appears in the definition of the LF current (the interaction $V(K)$ is present 
 both in the
projectors $\Pi(K)$ and in the 4D current ${\mathcal
J}$) and in the rhs of the LF WTI, where only 
$[g^{(n)}(K_f)]^{-1}$ and $[g^{(n)}(K_i)]^{-1}$ are 
affected by the interaction, as in the exact case, i.e. Eq. (\ref{wti3d}).

For the truncation at the n-th order, a truncated current that satisfies 
a truncated LF WTI is
given by \cite{MFPSS}
\be
j^{c (n)\mu}=j^{c (n-1)\mu} +\overline\Pi_0
\left[
\sum_{i=0}^{n}W_i\Delta_0\mathcal{J}_0^\mu\Delta_0W_{n-i}
+\sum_{i=0}^{n-1}W_i\Delta_0\mathcal{J}_I^\mu\Delta_0W_{n-1-i} \right]
\Pi_0~
\ee
where $W_i=V \left[\Delta_0 V\right]^{i-1}$ and it has been formally defined 
$W_0\Delta_0=\Delta_0W_0= 1$.
It should be reminded that $\mathcal{J}_0^\mu$ is $O(V^0)$ and $\mathcal{J}_I^\mu$ is $O(V^1)$.

Then $j^{c (n)\mu}$
satisfies a LF WTI given by
\be
Q^\mu{j_\mu^{c (n)}}= [g^{(n)}]^{-1}(K_f)~\widehat{\mathcal
Q}^L_{LF}-\widehat{\mathcal Q}^R_{LF}~[g^{(n)}]^{-1}(K_i)~~~. 
\ee

 For obtaining CC, the matrix
elements  should be taken between solutions
of $[g^{(n)}]^{-1}|\phi_n\rangle=0$.

\section{Two-body current in the LF approach: a pedagogical example}

A Yukawa model in
ladder approximation has been adopted in order to elaborate an application of
our general approach.
Within such a model, a couple of fermions interacts by the exchange of a boson,
in the present case chargeless, for simplicity. From the Lagrangian, one can
deduce immediately the 4D interaction to be used in the definition of the
T-matrix and of the 4D current. In ladder approximation, the 4D current is nothing
else but the free 4D current that fulfills the suitable WTI (see, e.g.
\cite{gross}). Let us stress that, though in  Minkowski
space the current of the model under investigation is the free one, its projection
onto a 3D hyperplane becomes interaction dependent. The matrix elements of the
first-order LF current, in a 3D Fourier space, are given by 
\be
\langle k_1^{\prime +}\vec{k}^\prime_{1\perp }|j^{c(1)\mu}|k_1^{+}\vec{k}_{1\perp }\rangle=
\langle k_1^{\prime +}\vec{k}^\prime_{1\perp }|\left [j^{c(0)\mu} +\overline
\Pi_0\left[V~\Delta_0\mathcal{J}^\mu_0 +\mathcal{J}^\mu_0 \Delta_0~V
\right]\Pi_0 \right] |k_1^{+}\vec{k}_{1\perp }\rangle\label{cvca1}
\ee
with $j^{c(0)\mu}=\overline
\Pi_0 \mathcal{J}_0^\mu
 \Pi_0$ (remind that $\Pi_0$ is the free-case projector).
In figure
1,  the diagrammatic analysis of
the first-order LF current is  shown. It should be pointed out that i) 
the diagrams are LF-time ordered, ii)
the LF-time flows from the right to the left and iii) the external legs 
are particles on their-own mass shell (therefore only three components can
independently change).
 Such a  diagrammatic analysis  
  allows one to appreciate the physical content
inside the operator $j^{c(1)}_\mu$. The peculiar feature of the LF
approach is the presence of instantaneous contributions, generated by the
term proportional to $\gamma^+$ in Eq. (\ref{proj2}). This contributions are labeled by horizontal
dashes. Another interesting contribution is  the
Z-diagram, or pair production diagram, generated by the time-ordering of the 4D
Dirac propagator. In this respect, we should emphasize that the full Dirac
structure, i.e. the spinorial structure, is exactly taken into account, since
the LF projection only works on the minus components of the four-momenta.

\begin{figure}
\includegraphics[width=3.7cm] {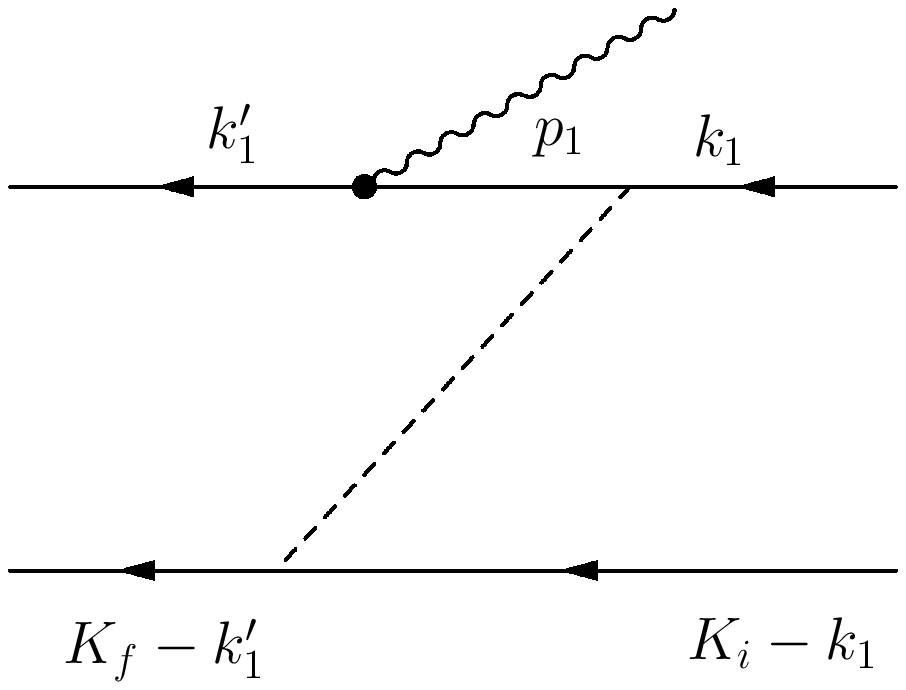}$~~~$
\includegraphics[width=3.7cm] {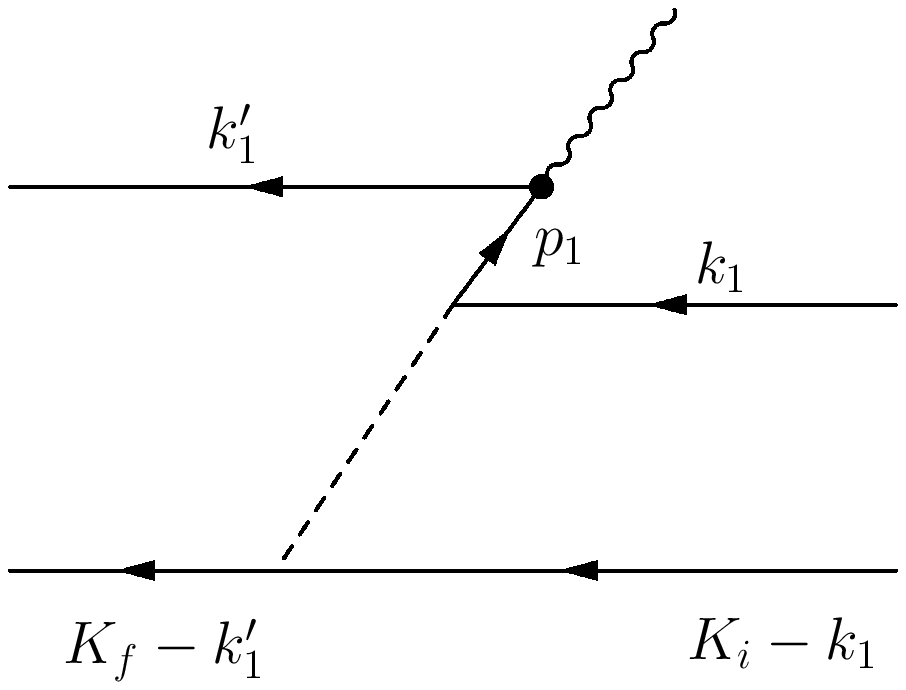}$~~~$
\includegraphics[width=3.7cm] {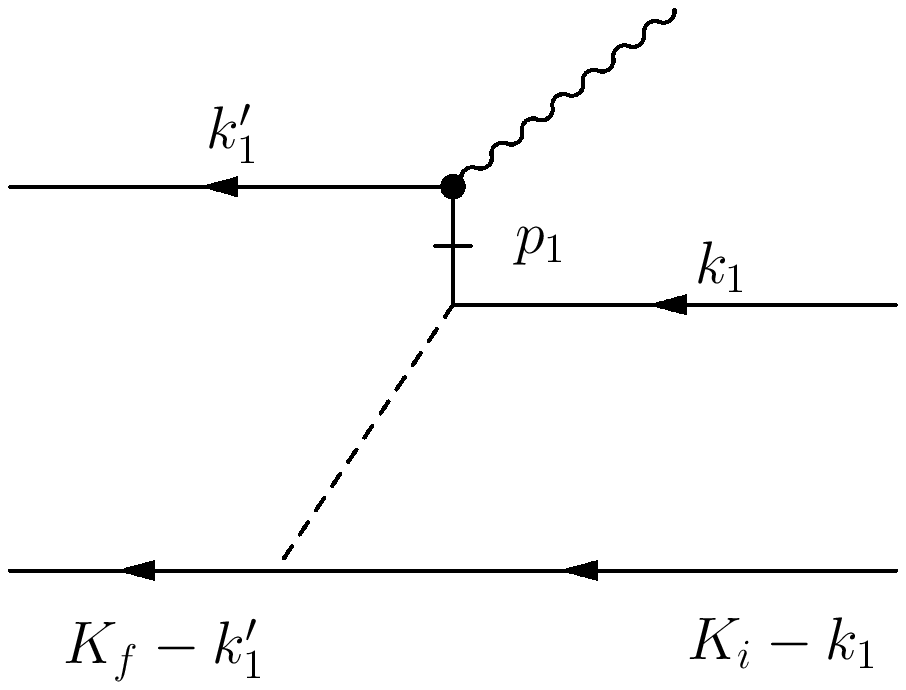}$~~~$
\includegraphics[width=3.7cm] {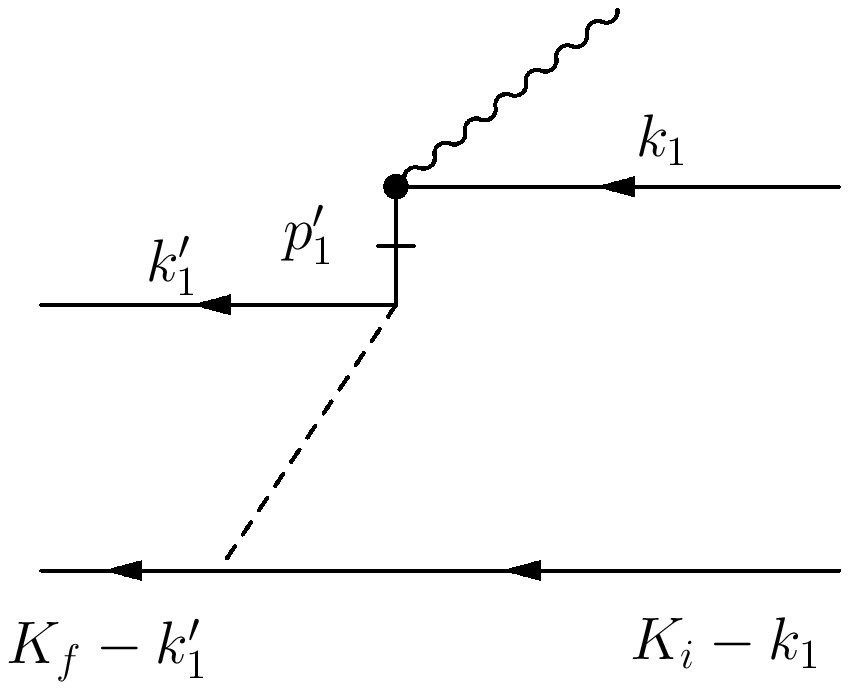}

$~~~~$

$~~~~$

\includegraphics[width=3.7cm] {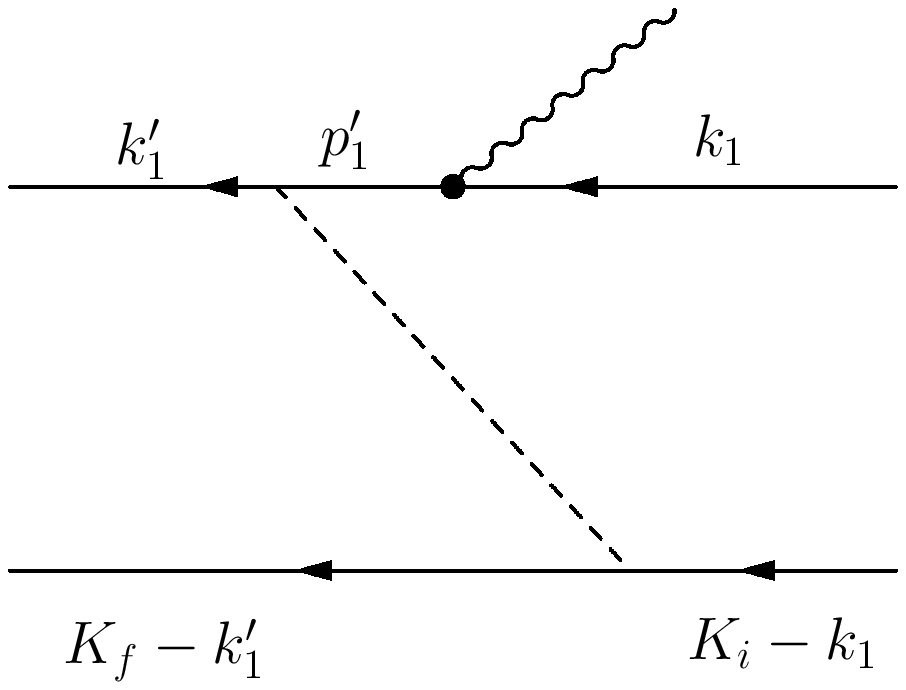}$~~~$
\includegraphics[width=3.7cm] {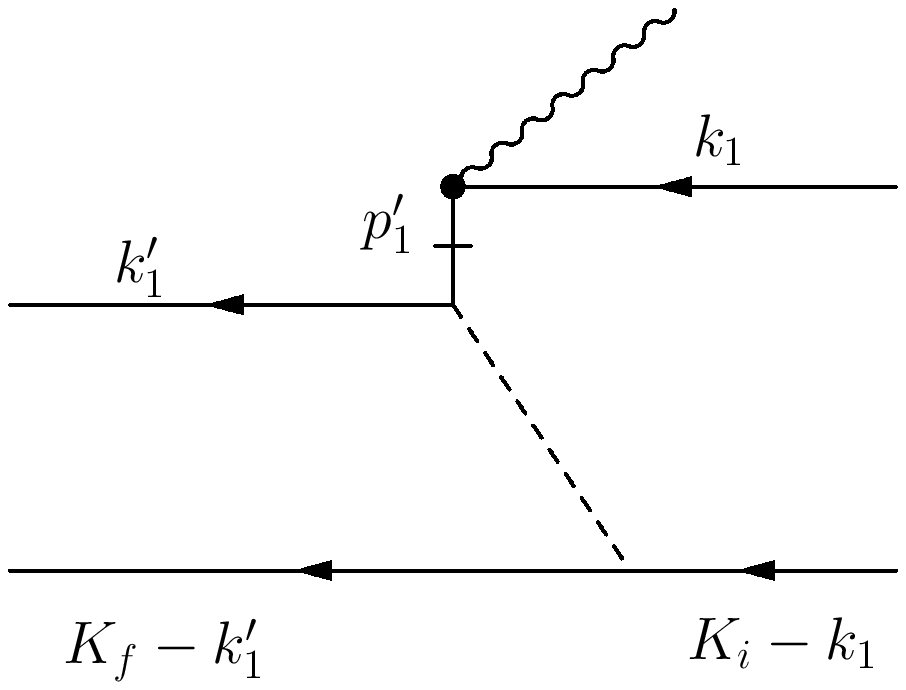}$~~~$
\includegraphics[width=3.7cm] {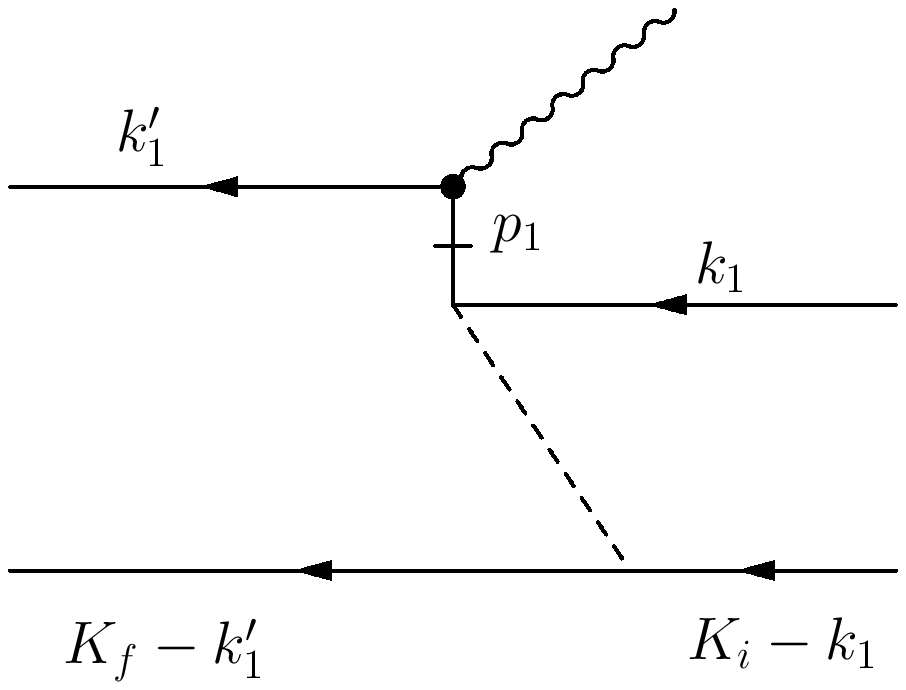}

\label{tbc}\caption{LF-time ordered diagrams for the first-order LF current for
a Yukawa model in the ladder approximation, see Eq. (\ref{cvca1}) and text.
Horizontal dashes indicate  instantaneous propagations.
(Adapted
from \cite{MFPSS})} \end{figure}

This  first-order current will be applied for evaluating the Deuteron
electromagnetic properties, in order to extend the analysis at the zero-order
already performed \cite{LPS00}. But,
one can
anticipate that i) the pair term  affects all the three electromagnetic form factors,
while the instantaneous terms  
contribute only to the magnetic one (since the property 
$\gamma^+ ~ \gamma^+=0$ is acting in the charge form factor), 
ii) the pair term vanishes for $q^+ \to
0$, for the conservation law of the plus components, while the 
instantaneous ones survive, iii) the pair term
should be maximal at $q^+\sim m_N$ \cite{NPA}, iv) 
the remaining terms
 affect all the Deuteron form factors, in the whole range of $q^+$. 

\section{Conclusions } 
For two interacting fermions, an exact correspondence between 4D BS 
amplitudes and 3D LF valence wave
functions has been obtained \cite{MFPSS}, by using the projection onto the LF hyperplane. 
The  
approach fulfills the covariance with respect to the kinematical 
Poincar\'e subgroup (7 generators out of 10).

The obtained relation allows one to express the matrix elements 
of the 4D
em current in terms of matrix elements of the 3D LF em current between valence
wave functions. For such a current operator, the corresponding LF WTI can be constructed by
introducing interaction free LF charge operators, then 
current conservation can be trivially retrieved.

Finally, by using a suitable auxiliary Green's
function, within the Quasi-Potential approach,
 that allows an ordering in terms of intermediate Fock states, an
approximation scheme has been developed, for both dynamical equation  and 
the em current operator. In particular, at any order of the effective 
interaction
the em current operator fulfills a proper LF WTI, leading to CC.

A systematic analysis of LF  two-body currents, obtained
within a Yukawa model in ladder approximation, 
 is
in progress for the Deuteron case.


\section*{References}
\begin{thebibliography}{99}
\bibitem{MFPSS} Marinho  J A O,  Frederico T,  Pace E,  
 Salm\`e G and  Sauer PU, 2008 {\it Phys. Rev. } D {\bf  77}  116010 
\bibitem{sales00}  Sales J H O,   Frederico T,   Carlson B V and  
Sauer P U, 2000 {\it Phys. Rev. } C {\bf   61}  044003  

\bibitem{sales01} Sales J H O,   Frederico .,   Carlson B V and  
Sauer P U, 2001 {\it Phys. Rev. } C {\bf   63}  064003
\bibitem{adnei07}  Marinho J A O,   Frederico T and  Sauer P U,
2007 {\it Phys. Rev. } D {\bf  76} 096001, and references quoted therein
 \bibitem{carb} Karmanov V A and   Carbonell J, 2006 {\it Eur. Phys.
J. } A {\bf  27}  1; 2006 {\it  Eur. Phys. J. } A {\bf  27} 11
\bibitem{grossbook}  Gross F, "Relativistic Quantum Mechanics and
Field Theory" (John Wiley \& Sons, New York 1993)
\bibitem{wolja}    Woloshyn R M and   Jackson A D, 1973 
{\it Nucl. Phys. } B {\bf  64} 269


\bibitem{brodsky}   Brodsky S J,   Pauli H C  and   Pinsky S S,
1998 {\it Phys. Rep.} {\bf 301}  299 
\bibitem{gross}   Gross F and   Riska D O, 1987 {\it Phys. Rev. } C {\bf   36} 
1928  
\bibitem{LPS00}   Lev F M,   Pace E and   Salm\`e G,  2000
{\it Phys. Rev. } C {\bf   62} 
   0640004 
\bibitem{NPA} De Melo J P B C,  Frederico T,  Pace E and    
 Salm\`e G, 2002 {\it Nucl. Phys. } A {\bf 707} 399
\end{thebibliography}
\end{document}